\documentclass[a4,12pt]{article}
\newcommand{\nc}{\newcommand}
\nc{\bg}{B. Grzadkowski}
\nc{\non}{\nonumber}
\def\dps{\displaystyle}
\def\mib#1{\mbox{\boldmath $#1$}}

\def\slak{\mbox{$k\!\!\! {/}$}\,}
\def\bra#1{\langle #1 |} \def\ket#1{|#1\rangle}
\def\vev#1{\langle #1\rangle}

\nc{\barx}{\bar{x}}\nc{\pbarn}{\;\hbox {pb}}
\nc{\fbarn}{\;\hbox {fb}} \nc{\hc}{\hbox {h.c.}}
\nc{\re}{\hbox {Re}} 
\nc{\mev}{\hbox {MeV}} \nc{\gev}{\;\hbox {GeV}}
\def\gesim{\lower0.5ex\hbox{$\:\buildrel >\over\sim\:$}}
\def\lesim{\lower0.5ex\hbox{$\:\buildrel <\over\sim\:$}}
\nc{\prd}[3]{{\it Phys.\ Rev.}\ {{\bf D{#1}} (#2) #3}}
\nc{\prl}[3]{{\it Phys.\ Rev.\ Lett.}\ {{\bf {#1}} (#2), #3}}
\nc{\plb}[3]{{\it Phys.\ Lett.}\ {{\bf B{#1}} (#2) #3}}
\nc{\npb}[3]{{\it Nucl.\ Phys.}\ {{\bf B{#1}} (#2) #3}}
\nc{\ptp}[3]{{\it Prog.\ Theor.\ Phys.}\ {{\bf {#1}} (#2), #3}}
\nc{\zfp}[3]{{\it Z.\ Phys.}\ {{\bf C{#1}} (#2) #3}}
\nc{\epj}[3]{{\it Eur.\ Phys.\ J.}\ {{\bf C{#1}} (#2) #3}}
\nc{\mpla}[3]{{\it Mod.\ Phys.\ Lett.}\ {{\bf A{#1}} (#2) #3}}
\nc{\rmp}[3]{{\it Rev.\ Mod.\ Phys.}\ {{\bf {#1}} (#2) #3}}
\nc{\ijmpa}[3]{{\it Int.\ J.\ Mod.\ Phys.}\
 {{\bf A{#1}} (#2) #3}}
\nc{\ttbar}{t\bar{t}}  \nc{\bbbar}{b\bar{b}}
\nc{\tanb}{\tan \beta}  \nc{\twbdec}{t\to W^+ b}
\nc{\tbwbdec}{\bar{t}\to W^- \bar{b}}
\nc{\epem}{e^+e^-}  \nc{\eett}{\epem \to \ttbar}
\nc{\sigeett}{\sigma_{e\bar{e}\to\ttbar}}
\nc{\wpwm}{W^+W^-}  \nc{\tbar}{\bar{t}}
\nc{\bbar}{\bar{b}}  \nc{\wpp}{W^+}
\nc{\mt}{m_t}  \nc{\mts}{m_t^2}  \nc{\mw}{m_W}  \nc{\mws}{m_W^2}
\nc{\mz}{m_Z}  \nc{\mzs}{m_Z^2}
\nc{\ttbardec}{\ttbar \to W^+W^-\bbbar}
\nc{\wwbb}{W^+W^-\bbbar}  \nc{\sm}{SM}
\nc{\cw}{\cos\theta_W}  \nc{\sw}{\sin\theta_W}
\nc{\sws}{\sin^2\theta_W}  \nc{\sig}{\sigma_{tot}}
\nc{\lp}{{\ell}^+}  \nc{\lm}{{\ell}^-}
\nc{\epsl}{\epsilon_L}  \nc{\cp}{C\!P}
\nc{\gaga}{\gamma\gamma}
\nc{\splus}{s_+}  \nc{\smin}{s_-}  \nc{\eps}{\epsilon}
\nc{\psp}{Ps_+}  \nc{\psm}{Ps_-}  \nc{\lsp}{ls_+}
\nc{\lsm}{ls_-}  \nc{\sss}{s_+s_-}  \nc{\m}{m_t}
\nc{\mq}{m_t^2}  \nc{\mr}{\frac{1}{\m}} \nc{\av}{A_{\gamma}}
\nc{\bv}{B_{\gamma}}  \nc{\az}{A_Z}  \nc{\bz}{B_Z}
\nc{\avs}{A_{\gamma}^2}\nc{\azs}{A_Z^2}  \nc{\bzs}{B_Z^2}
\nc{\dav}{\delta \! A_{\gamma}}  \nc{\dbv}{\delta \! B_{\gamma}}
\nc{\dcv}{\delta C_{\gamma}}  \nc{\ddv}{\delta \! D_{\gamma}}
\nc{\daz}{\delta \! A_Z}  \nc{\dbz}{\delta \! B_Z}
\nc{\dcz}{\delta C_Z}  \nc{\ddz}{\delta \! D_Z}
\nc{\dev}{\delta \! E_{\gamma}}  \nc{\dez}{\delta \! E_Z}
\nc{\dfv}{\delta \! F_{\gamma}}  \nc{\dfz}{\delta \! F_Z}
\nc{\rdav}{{\rm Re}(\delta \! A_{\gamma}) \:}
\nc{\rdbv}{{\rm Re}(\delta \! B_{\gamma}) \:}
\nc{\rdcv}{{\rm Re}(\delta C_{\gamma}) \:}
\nc{\rddv}{{\rm Re}(\delta \! D_{\gamma}) \:}
\nc{\rdaz}{{\rm Re}(\delta \! A_Z) \:}
\nc{\rdbz}{{\rm Re}(\delta \! B_Z) \:}
\nc{\rdcz}{{\rm Re}(\delta C_Z) \:}
\nc{\rddz}{{\rm Re}(\delta \! D_Z) \:}
\nc{\idav}{{\rm Im}(\delta \! A_{\gamma}) \:}
\nc{\idbv}{{\rm Im}(\delta \! B_{\gamma}) \:}
\nc{\idcv}{{\rm Im}(\delta C_{\gamma}) \:}
\nc{\iddv}{{\rm Im}(\delta \! D_{\gamma}) \:}
\nc{\idaz}{{\rm Im}(\delta \! A_Z) \:}
\nc{\idbz}{{\rm Im}(\delta \! B_Z) \:}
\nc{\idcz}{{\rm Im}(\delta C_Z) \:}
\nc{\iddz}{{\rm Im}(\delta \! D_Z) \:}
\nc{\cz}{(1+v_e^2)d\:\!'^2}  \nc{\ci}{v_ed\:\!'}
\nc{\ccz}{v_ed\:\!'^2}  \nc{\cci}{d\:\!'}
\nc{\lspace}{\;\;\;\;\;\;\;\;\;\;}  \nc{\llspace}{\lspace \lspace}
\nc{\beq}{\begin{equation}}  \nc{\eeq}{\end{equation}}
\nc{\bea}{\begin{eqnarray}}  \nc{\eea}{\end{eqnarray}}
\nc{\baa}{\begin{array}}  \nc{\eaa}{\end{array}}
\nc{\bit}{\begin{itemize}}  \nc{\eit}{\end{itemize}}
\nc{\ben}{\begin{enumerate}}  \nc{\een}{\end{enumerate}}
\nc{\bce}{\begin{center}}  \nc{\ece}{\end{center}}
\nc{\ocal}{{\cal O}}
\newcounter{QQ}
\newcounter{QQF}
\newcounter{QQE}
\begin{document}
\pagestyle{empty} \setlength{\footskip}{2.0cm}
\setlength{\oddsidemargin}{0.5cm}
\setlength{\evensidemargin}{0.5cm}
\renewcommand{\thepage}{-- \arabic{page} --}
\def\mib#1{\mbox{\boldmath $#1$}}
\def\bra#1{\langle #1 |}  \def\ket#1{|#1\rangle}
\def\vev#1{\langle #1\rangle} \def\dps{\displaystyle}
\nc{\tb}{\stackrel{{\scriptscriptstyle (-)}}{t}}
\nc{\bb}{\stackrel{{\scriptscriptstyle (-)}}{b}}
\nc{\fb}{\stackrel{{\scriptscriptstyle (-)}}{f}}
\nc{\pp}{\gamma \gamma}
\nc{\pptt}{\pp \to \ttbar}
 \def\thebibliography#1{\centerline{REFERENCES}
 \list{[\arabic{enumi}]}{\settowidth\labelwidth{[#1]}\leftmargin
 \labelwidth\advance\leftmargin\labelsep\usecounter{enumi}}
 \def\newblock{\hskip .11em plus .33em minus -.07em}\sloppy
 \clubpenalty4000\widowpenalty4000\sfcode`\.=1000\relax}\let
 \endthebibliography=\endlist
 \def\sec#1{\addtocounter{section}{1}\section*{\hspace*{-0.72cm}
 \normalsize\bf\arabic{section}.$\;$#1}\vspace*{-0.3cm}}
\vspace*{-1.7cm}
\noindent
 \phantom{\large\bf Modified : September 28, 2005}

 \vspace{-0.7cm}
\begin{flushright}
$\vcenter{
{\hbox{{\footnotesize IFT-20-05~~~~FUT-05-01}}}
{\hbox{{\footnotesize UCRHEP-T396}}}
{\hbox{{\footnotesize TOKUSHIMA Report}}}
{\hbox{(hep-ph/0508183)}}
}$
\end{flushright}

\vskip 0.6cm
\begin{center}
{\large\bf Optimal Beam Polarizations for New-Physics Search}

\vskip 0.15cm
{\large\bf through $\mib{\gamma\gamma\to t\bar{t}\to \ell X/bX}$}
\end{center}

\vspace{0.2cm}
\begin{center}
\renewcommand{\thefootnote}{\alph{footnote})}
{\sc Bohdan GRZADKOWSKI$^{\:1),\:}$}\footnote{E-mail address:
\tt bohdan.grzadkowski@fuw.edu.pl},\ \
{\sc Zenr\=o HIOKI$^{\:2),\:}$}\footnote{E-mail address:
\tt hioki@ias.tokushima-u.ac.jp},

\vskip 0.15cm
{\sc Kazumasa OHKUMA$^{\:3),\:}$}\footnote{E-mail address:
\tt ohkuma@fukui-ut.ac.jp}\ and\
{\sc Jos\'e WUDKA$^{\:4),\:}$}\footnote{E-mail address:
\tt jose.wudka@ucr.edu}
\end{center}

\vspace*{0.2cm}
\centerline{\sl $1)$ Institute of Theoretical Physics,\ Warsaw
University}
\centerline{\sl Ho\.za 69, PL-00-681 Warsaw, Poland}

\vskip 0.2cm
\centerline{\sl $2)$ Institute of Theoretical Physics,\
University of Tokushima}
\centerline{\sl Tokushima 770-8502, Japan}

\vskip 0.2cm
\centerline{\sl $3)$ Department of Information Science,\
Fukui University of Technology}
\centerline{\sl Fukui 910-8505, Japan}

\vskip 0.2cm
\centerline{\sl $4)$ Department of Physics,\
University of California, Riverside}
\centerline{\sl Riverside, CA 92521-0413, USA}

\vspace*{1.2cm}
\centerline{ABSTRACT}

\vspace*{0.3cm}
\baselineskip=20pt plus 0.1pt minus 0.1pt
We perform an optimal-observable analysis of the final
charged-lepton/$b$-quark momentum distributions in
$\gamma\gamma\to t\tbar \to \ell X/bX$ for various beam
polarizations in order to study possible anomalous
$t\bar{t}\gamma$, $tbW$ and $\gamma\gamma H$ couplings,
which could be generated by $SU(2)\times U(1)$
gauge-in\-var\-i\-ant dimension-6 effective operators.
We find optimal beam polarizations that will minimize the
uncertainty in determination of those non-standard couplings.
We also compare $e\bar{e}$ and $\gamma\gamma$ colliders
from the viewpoint of the anomalous-top-quark-coupling
determination.
\vspace*{0.4cm} \vfill

PACS:  14.65.Fy, 14.65.Ha, 14.70.Bh

Keywords:
anomalous top-quark couplings, $\gamma\gamma$ colliders \\

\newpage
\renewcommand{\thefootnote}{$\sharp$\arabic{footnote}}
\pagestyle{plain} \setcounter{footnote}{0}
\baselineskip=20.5pt plus 0.2pt minus 0.1pt

\sec{Introduction}

Top-quark and Higgs-boson sectors are still not
fully-tested regions of the electroweak physics. If there
exists any new physics beyond the Standard Model (SM),
it is plausible that its effects appear in those sectors.
Therefore it is worth looking for experiments that are sensitive
to top-quark and Higgs-boson properties, in particular
to deviations from the SM predictions.
Anomalous top-quark interactions will be tested by the various
programs envisaged by the International Linear Collider (ILC)
project \cite{LP2005}. In particular the photon-photon
mode \cite{Ginzburg:1981vm}--\cite{Borden:1993cw} of this collider
will be able to probe efficiently the top-quark properties
through $\ttbar$ production, as well as the Higgs-boson interactions.
Therefore the $\gaga$ collider will prove to be a
useful tool for searching for non-standard physics.

Indeed, compared to $e\bar{e}$ machines, $\gamma\gamma$ colliders
present remarkable advantages, e.g., for the study of $C\!P$
violation in $\gaga H$ couplings \cite{Grzadkowski:1992sa}.
In the case of $e\bar{e}$ collisions, the only relevant
initial states are $C\!P$-even states
$\ket{e_{L/R}\bar{e}_{R/L}}$ under the usual assumption that
the electron mass can be neglected and that an $e\bar{e}$
pair annihilates dominantly into a single (virtual)
vector-/axial-vector-boson. Therefore, all
$C\!P$-violating observables must be constructed there from
final-particle momenta/polarizations. In contrast, a
$\gamma\gamma$ collider offers the unique possibility of
preparing the polarization of the incident-photon beams which
can be used to construct $C\!P$-violating asymmetries without
relying on final-state information \cite{Grzadkowski:1992sa}.

Because of this a number of authors have already considered top-quark
production and decays in $\gamma\gamma$ collisions in order to
study {\it i}) Higgs-boson couplings to the top quark and
photon \cite{Anlauf:1995mu}--\cite{Asakawa:2003dh}, or
{\it ii}) anomalous top-quark couplings to the photon
\cite{Choi:1995kp}--\cite{Poulose:1997xk}. However, what
will be observed in real experiments are combined signals that
originate both from the process of top-quark production and,
{\it in addition}, from its decays. Therefore, we have
recently performed
a model-independent analysis of $\gamma\gamma \to \ttbar \to
{\ell} X/bX$ \cite{Grzadkowski:2003tf},
including all possible non-standard interactions together
(production and decay) and applying the optimal-observable
(OO) procedure to the final charged-lepton/$b$-quark momentum
distributions.

In this work, we present a comprehensive
analysis based on that framework, aiming to find optimal beam
polarizations that minimize the uncertainty in determination of
$t\bar{t}\gamma$- , $tbW$- and $\gamma\gamma H$-coupling parameters.
Concerning the Higgs couplings,
we do not intend to go into its resonance region since our main
interest is in $t\bar{t}$ production/decay and also that region
has already been studied in great details in existing
literature \cite{Grzadkowski:1992sa}-\cite{Asakawa:2003dh}.
Another goal here is to compare the $e\bar{e}$
and $\gamma\gamma$ colliders from the view point of the
anomalous-top-quark-coupling determination.

The outline of this paper is
as follows. After summarizing our fundamental framework in
sec. 2, we give detailed numerical results of the analysis
in sec. 3. In sec. 4 we perform a comparison of the $e\bar{e}$
and $\gamma\gamma$ colliders and the final section is devoted
to conclusions and discussions.
Some basic formulas and tools used in the analysis are
described in detail in the appendix.

\sec{Framework}

In this section, we summarize the basic elements of the
framework used in the analyses, parts of which are
described in more detail in appendices A1 and A2.

\paragraph{\bf Effective Lagrangian}

We have used an effective low-energy Lagrangian parameterization
\cite{Buchmuller:1986jz} in order to describe possible
new-physics effects, i.e., the SM Lagrangian is modified by
the addition of a series of $SU(3)\times SU(2)\times U(1)$
gauge-invariant operators, which are suppressed by inverse
powers of a new-physics scale ${\mit\Lambda}$. Among those
operators, the largest contribution comes from dimension-6
operators,\footnote{Dimension-5 operators are not included
    since they violate lepton number \cite{Buchmuller:1986jz}
    and are irrelevant for the processes considered here.}
denoted as ${\cal O}_i$, and we have the effective Lagrangian as
\begin{equation}
{\cal L}_{\rm eff}={\cal L}_{\rm SM}
+\frac1{{\mit\Lambda}^2}\sum_i (\alpha_i {\cal O}_i
+ {\rm h.c.}) + O( {\mit\Lambda}^{-3} ).
\end{equation}
The operators relevant here lead to the
following non-standard top-quark- and Higgs-boson-couplings:
(1) $C\!P$-conserving and $C\!P$-violating $t\bar{t}\gamma$ vertices,
(2) $C\!P$-conserving and $C\!P$-violating $\gamma\gamma H$ vertices, and
(3) the anomalous $tbW$ vertex. The corresponding coupling
constants are denoted respectively by
the five independent parameters
$\alpha_{\gamma 1}$, $\alpha_{\gamma 2}$, $\alpha_{h1}$,
$\alpha_{h2}$ and $\alpha_{d}$.
The explicit expressions for
these couplings in terms of the coefficients
of dimension-6 operators are to be found in appendices A1 and A2.

It is worth pointing out that the effective-Lagrangian
parameterization is equally applicable to
$e\bar{e}\to t\bar{t}$
and $\gamma\gamma\to t\bar{t}$. In the former case,
no additional complications are encountered when
replacing the effective-operator vertices by form factors
(as given in appendix A3; see also
\cite{Grzadkowski:1996kn,Grzadkowski:2000nx}) since
all kinematic variables are fixed
by the CM energy $\sqrt{s}$. This situation does not recur
in $\gamma\gamma\to t\bar{t} $: the kinematic variables in
the $t$-channel top exchange are not fixed
by $s$, so, if we replace the effective couplings by
form factors, the cross section will
depend on the functional form of the latter. We
will return to this point below.

\paragraph{\bf ${\mib \gamma}{\mib \gamma}$ colliders}

Following the standard approach \cite{Ginzburg:1981vm}, each
photon beam originates as a laser beam back-scattered off an
electron ($e$) or positron ($\bar{e}$) beam. The polarizations
of the initial-state are  characterized by the
electron and positron longitudinal polarizations $P_e$
and $P_{\bar{e}}$, the maximum average linear polarizations
$P_t$ and $P_{\tilde{t}}$ of the laser photons with
the azimuthal angles $\varphi$ and $\tilde{\varphi}$
(defined in the same way as in \cite{Ginzburg:1981vm}), and
their average helicities $P_{\gamma}$ and $P_{\tilde{\gamma}}$.
The photon polarizations $P_{t,\gamma}$ and
$P_{\tilde{t},\tilde{\gamma}}$ satisfy
\begin{equation}
0 \leq P_t^2 + P_{\gamma}^2 \leq 1,
\ \ \ \ \ \ \ \ \
0 \leq P_{\tilde{t}}^2 + P_{\tilde{\gamma}}^2 \leq 1,
\end{equation}
and combine  with the azimuthal angles to
form the following polarization density matrices:
\begin{equation}
\rho=\frac12
\left(\begin{array}{cc}1+P_{\gamma} & -P_t e^{-2i\varphi}\ \\
-P_t e^{2i\varphi} & 1-P_{\gamma} \end{array}\right),
\ \ \ \
\tilde{\rho}=\frac12
\left(\begin{array}{cc}1+P_{\tilde{\gamma}} &
-P_{\tilde{t}} e^{2i\tilde{\varphi}}\ \\
-P_{\tilde{t}} e^{-2i\tilde{\varphi}} &
1-P_{\tilde{\gamma}} \end{array}\right).
\end{equation}

For  linear polarization, we denote the relative azimuthal
angle by $\chi \equiv \varphi-\tilde{\varphi}$, which we
fixed to be $\pi/4$ by the following procedure: we calculated
the cross section $\sigma(\gamma\gamma\to t\bar{t})$
to first order in $\alpha_{\gamma 1,\gamma2,h1,h2}$, we
found that the terms proportional to $\alpha_{\gamma2}$ and
those to $\alpha_{h2}$ were the most sensitive to $\chi$ and
that these were maximized when $\chi=\pi/4$
(this was previously noticed in
\cite{Choi:1995kp} concerning the $\alpha_{\gamma 2}$ term).

\paragraph{\bf Cross sections}

When calculating the cross section
$d\sigma(\gamma\gamma\to t\bar{t})$,
the photon beams do not have definite spins and momenta
as in $e\bar{e}/p\bar{p}$ colliders; for back scattered
photon the spin information is given in terms of
the Stokes parameters and the momentum distribution
by the photon spectrum function. The
cross section is calculated similarly to parton-model
calculations (see, for example,
\cite{Grzadkowski:2003tf} for more details). Taking this into account
the calculation is straightforward but the final
expressions are very lengthy and will not be
displayed here; to simplify the
algebraic manipulations we used FORM \cite{FORM}.

In deriving the distributions of secondary fermions $(=\ell/b)$
produced by the above cross section and the decay
widths $ d{\mit\Gamma}(t \to \ell X/bX)$, we use the narrow-width
approximation thus treating the
decaying $t$ and $W$ as on-shell particles; this
enables us to use the Kawasaki-Shirafuji-Tsai formalism
\cite{technique}. We have also neglected all contributions quadratic in
$\alpha_i$ ($i=\gamma 1,\gamma 2, h1,h2,d$), so that the
angular-energy distributions of the secondary fermions
$\ell/b$ in the $e\bar{e}$ (the initial electron-positron beams)
CM frame can be expressed as
\begin{equation}
\frac{d\sigma}{dE_{\ell/b} d\cos\theta_{\ell/b}}
=f_{\rm SM}(E_{\ell/b}, \cos\theta_{\ell/b})
 + \sum_i \alpha_i f_i (E_{\ell/b}, \cos\theta_{\ell/b}),
\label{distribution}
\end{equation}
where $f_{\rm SM}$ and $f_i$ are calculable functions:
$f_{\rm SM}$ denotes the SM contribution,
$f_{\gamma 1,\gamma 2}$ describe the anomalous
$C\!P$-conserving and $C\!P$-violating
$t\bar{t}\gamma$-ver\-ti\-ces contributions respectively,
$f_{h1,h2}$ those generated by the anomalous $C\!P$-conserving
and $C\!P$-violating $\gamma\gamma H$-ver\-ti\-ces,
and $f_d$ that by the anomalous $tbW$-vertex.

\paragraph{\bf Optimal-observable technique}

The optimal-observable technique \cite{optimal} is a useful tool
for estimating expected statistical uncertainties in various
coupling measurements. Suppose we have a cross section
\begin{equation}
\frac{d\sigma}{d\phi}(\equiv{\mit\Sigma}(\phi))=\sum_i c_i f_i(\phi),
\end{equation}
where $f_i(\phi)$ are known functions of the location in final-state
phase space variables $\phi$ and $c_i$'s are model-dependent coefficients. 
The
goal is to determine the $c_i$'s. This can be done by using
appropriate weighting functions $w_i(\phi)$ such that $\int w_i(\phi)
{\mit\Sigma}(\phi)d\phi=c_i$. In general different choices for
$w_i(\phi)$ are possible, but there is a unique choice for which the
resultant statistical error is minimized. Such functions are given by
\begin{equation}
w_i(\phi)=\sum_j X_{ij}f_j(\phi)/{\mit\Sigma}(\phi)\,, \label{X_def}
\end{equation}
where $X_{ij}$ is the inverse matrix of ${\cal M}_{ij}$ which
is defined as
\begin{equation}
{\cal M}_{ij}
\equiv \int {f_i(\phi)f_j(\phi)\over{\mit\Sigma}(\phi)}d\phi\,.
\label{M_def}
\end{equation}
When we use these weighting functions, the statistical uncertainty
of $c_i$ becomes
\begin{equation}
{\mit\Delta}c_i=\sqrt{X_{ii}\,\sigma_T/N}\,, \label{delc_i}
\end{equation}
where $\sigma_T\equiv\int (d\sigma/d\phi) d\phi$ and $N$ is the total
number of events.

In order to apply this technique to eq.(\ref{distribution}),
we first have to calculate ${\cal M}_{ij}$ using $f_{\rm SM}$
and $f_i$
\begin{equation}
{\cal M}_{ij}=\int dE_{\ell/b} d\cos\theta_{\ell/b}\,
f_i(E_{\ell/b}, \cos\theta_{\ell/b})
f_j(E_{\ell/b}, \cos\theta_{\ell/b})/
f_{\rm SM}(E_{\ell/b}, \cos\theta_{\ell/b})
\end{equation}
and its inverse matrix $X_{ij}$, where $i,j=1,\cdots, 6$
correspond to SM, $\gamma 1$, $\gamma 2$, $h1$, $h2$ and $d$
respectively. Then, according to eq.(\ref{delc_i}), the expected
statistical uncertainty for the measurements of $\alpha_i$
is given by
\beq
{\mit\Delta}\alpha_i=\sqrt{X_{ii}\,\sigma_T/N_{\ell/b}},
\eeq
where
\beq
\,\sigma_T = \int dE_{\ell/b}
d\cos\theta_{\ell/b}\, f_{\rm SM}(E_{\ell/b}, \cos\theta_{\ell/b}).
\eeq

In this calculation we will not probe the Higgs-resonance region
which has been extensively studied previously (see, for example,
\cite{Gounaris:1997ef}). Therefore, since we work to lowest
order in the $\alpha_i$, we
compute the number of secondary fermions, $N_{\ell/b}$,
from the SM total cross section
multiplied by the lepton/$b$-quark detection efficiency
$\epsilon_{\ell/b}$ and the integrated $e\bar{e}$ luminosity
$L_{e\bar{e}}\,$; this leads to $N_{\ell/b}$ independent of
$m_H$.

\sec{Anomalous couplings and optimal polarizations}

In our previous work \cite{Grzadkowski:2003tf}, where our main
concern was to construct a fundamental framework for practical
analyses, we used (1) $P_e=P_{\bar{e}}=1$ and
$P_t =P_{\tilde{t}}=P_{\gamma}=P_{\tilde{\gamma}}=1/\sqrt{2}$,
and (2) $P_e=P_{\bar{e}}=P_{\gamma}=P_{\tilde{\gamma}}=1$ as
typical polarization examples and performed an OO-analysis.
Inverting the matrix ${\cal M}_{ij}$, we noticed that the
numerical results for $X_{ij}$ are often unstable
\cite{Grzadkowski:2003tf}: even a tiny variation of ${\cal M}_{ij}$
changes $X_{ij}$ significantly. This indicates that some of
$f_i$ have similar shapes and therefore their coefficients
cannot be disentangled easily.
The presence of such instability forced us to forgo our initial goal of
determining all the couplings at once through this process alone.
That is, we assume that some of $\alpha_i$'s have been
measured in other processes (e.g., in
$e\bar{e}\to t\bar{t}\to{\ell}^{\pm}X$), and we performed an
analysis with smaller number of independent parameters.

For example, when estimating the statistical uncertainty in simultaneous
measurements of $\alpha_{\gamma 1}$ and $\alpha_{h 1}$
(assuming all other coefficients are known), we need only
the matrix components with indices 1, 2 and 4. In such a ``reduced
analysis'', we allowed only ``stable solutions" according to the following
criterion: we calculate the selected ${\mit\Delta}\alpha_i$ rounding
${\cal M}_{ij}$
first to three and then to two decimal places, obtaining
${\mit\Delta}\alpha_i^{[3]}$ and ${\mit\Delta}\alpha_i^{[2]}$ respectively.
We then accept the result as a
stable solution if
$|{\mit\Delta}\alpha_i^{[3]}
-{\mit\Delta}\alpha_i^{[2]}|/
{\mit\Delta}\alpha_i^{[3]} \leq 0.1$ for $i=\gamma1,~h1$.

In this work, we took $\sqrt{s_{e\bar{e}}}=500$ GeV and ${\mit\Lambda}=1$
TeV and minimized the statistical uncertainties ${\mit\Delta} \alpha_i$
by choosing the polarization parameters from
the set $\{P_{e,\bar{e}}=0,\,\pm1; ~ P_{t,\tilde{t}}=0,
\,1/\sqrt{2},\,1; ~P_{\gamma,\tilde{\gamma}}=0,
\,\pm 1/\sqrt{2},\,\pm1 \}$.
We also considered 3 values of the Higgs mass, $m_H=100,~ 300$ and $500$ 
GeV,
which correspond to the widths ${\mit\Gamma}_H=1.08\times 10^{-2},~ 8.38$
and $73.4$ GeV respectively according to the standard-model formula.

Although we again did not find any stable solution in the four-
and five-parameter analysis, we did find some solutions not only
in the two- but also in the three-parameter analysis. This is
in marked contrast to the results in \cite{Grzadkowski:2003tf},
where we found no stable solution for the three-parameter analysis.
In order to insure acceptable statistical precision we
required the solutions to satisfy the following conditions:
\begin{itemize}
\item {\sl Three-parameter analysis:} the resulting uncertainties must obey
       $ {\mit\Delta}\alpha_i \leq 0.1 $  for at least two of the
       three unknown couplings, for an integrated luminosity of
       $L_{e\bar{e}}=500\ {\rm fb}^{-1}$ (without
       detection-efficiency suppression) .
\item {\sl Two-parameter analysis:} after selecting the
Higgs-boson mass and the secondary
fermion ($b$ or $\ell $) that will be observed,
we found many stable
solutions. We then selected  those
pairs $\{ {\mit\Delta}\alpha_i , {\mit\Delta}\alpha_j \}$ that satisfy
${\mit\Delta}\alpha_{i,j}\leq 0.1$ for a luminosity of
$L_{e\bar{e}}=500\ {\rm fb}^{-1}$, and which minimize
$({\mit\Delta}\alpha_i)^2+({\mit\Delta}\alpha_j)^2$.
\end{itemize}

The results are presented below. We did not fix the detection
efficiencies $\epsilon_{\ell/b}$ since they depend
on detector parameters and will improve with the development of
detection technology.

\setcounter{QQ}{\arabic{equation}} \addtocounter{QQ}{1}
\begin{description}
\item[1)] Three parameter analysis
\item[] $\oplus$ Final charged-lepton detection
\begin{description}
  \item[] $m_{H}=500$ GeV
  \item[$\bullet$]
    $P_e=P_{\bar{e}}=0,~P_t = P_{\tilde{t}}=1/\sqrt{2},~
     P_\gamma = - P_{\tilde{\gamma}}=1/\sqrt{2}$,
     $N_{\ell}\simeq 6.1\times 10^3 \epsilon_{\ell}$\\
     ${\mit\Delta} \alpha_{\gamma2}=0.94/\sqrt{\epsilon_{\ell}},~
      {\mit\Delta} \alpha_{h2}=0.11/\sqrt{\epsilon_{\ell}} ,~
      {\mit\Delta} \alpha_{d}=0.042/\sqrt{\epsilon_{\ell}}$.
\hfill (\arabic{QQ})

Strictly speaking, this result does not satisfy our condition
for the three-parameter analysis, but we show it since
${\mit\Delta} \alpha_{h2}$ exceeds the limit by only 0.01. 
\setcounter{QQF}{\arabic{QQ}} \addtocounter{QQ}{1}
\end{description}
\item[] $\oplus$ Final bottom-quark detection
\begin{description}
  \item[] $m_{H}=100$ GeV
  \item[$\bullet$]
    $P_e=P_{\bar{e}}=1,~P_t = P_{\tilde{t}}=1/\sqrt{2},~
     P_\gamma = - P_{\tilde{\gamma}}=1/\sqrt{2}$, \ \
     $N_{b}\simeq 4.2\times 10^4 \epsilon_{b}$\\
     ${\mit\Delta} \alpha_{h1}=0.086/\sqrt{\epsilon_{b}}$,~
     ${\mit\Delta} \alpha_{h2}=0.21/\sqrt{\epsilon_{b}}$,~
     ${\mit\Delta} \alpha_{d}=0.037/\sqrt{\epsilon_{b}}$.
\hfill (\arabic{QQ})\addtocounter{QQ}{1}
  \item[] $m_{H}=500$ GeV
  \item[$\bullet$]
    $P_e=P_{\bar{e}}=0,~P_t = P_{\tilde{t}}=1/\sqrt{2},~
     P_\gamma = - P_{\tilde{\gamma}}=1/\sqrt{2}$, \ \
     $N_{b}\simeq 2.8\times 10^4 \epsilon_{b}$\\
     ${\mit\Delta} \alpha_{\gamma2}=0.61/\sqrt{\epsilon_{b}}$,~
     ${\mit\Delta} \alpha_{h2}=0.054/\sqrt{\epsilon_{b}}$,~
     ${\mit\Delta} \alpha_{d}=0.052/\sqrt{\epsilon_{b}}$.
\hfill (\arabic{QQ})\addtocounter{QQ}{1}
 \end{description}
\item[2)] Two parameter analysis
\item[] $\oplus$ Final charged-lepton detection
\begin{description}
    \item[] Independent of $m_H$
  \item[$\bullet$]
    $P_e=P_{\bar{e}}=-1,~P_t = P_{\tilde{t}}=1,~
     P_\gamma = P_{\tilde{\gamma}}=0$, \ \
     $N_{\ell}\simeq 1.0\times 10^4 \epsilon_{\ell}$\\
     ${\mit\Delta} \alpha_{\gamma1}=0.051/\sqrt{\epsilon_{\ell}}$,~
     ${\mit\Delta} \alpha_{d}=0.022/\sqrt{\epsilon_{\ell}}$.
\hfill (\arabic{QQ})\addtocounter{QQ}{1}

This result is independent of  $m_H$  since the Higgs-exchange
diagram does not contribute to the determination of
$\alpha_{\gamma1}$ and $\alpha_d$ within our approximation.
  \item[] $m_H=100$ GeV
  \item[$\bullet$]
    $P_e=P_{\bar{e}}=-1,~P_t = P_{\tilde{t}}=1/\sqrt{2},~
     P_\gamma = P_{\tilde{\gamma}}=1/\sqrt{2}$, \ \
     $N_{\ell}\simeq 1.9\times 10^4 \epsilon_{\ell}$\\
     ${\mit\Delta} \alpha_{h1}=0.034/\sqrt{\epsilon_{\ell}}$,~
     ${\mit\Delta} \alpha_{d}=0.017/\sqrt{\epsilon_{\ell}}$.
\hfill (\arabic{QQ})\addtocounter{QQ}{1}
  \item[] $m_H=300$ GeV
  \item[$\bullet$]
    $P_e=P_{\bar{e}}=-1,~P_t = P_{\tilde{t}}=0,~
     P_\gamma = P_{\tilde{\gamma}}=1$, \ \
     $N_{\ell}\simeq 2.4\times 10^4 \epsilon_{\ell}$\\
     ${\mit\Delta} \alpha_{h1}=0.013/\sqrt{\epsilon_{\ell}}$,~
     ${\mit\Delta} \alpha_{d}=0.015/\sqrt{\epsilon_{\ell}}$.
\hfill (\arabic{QQ})\addtocounter{QQ}{1}
  \item[] $m_H=500$ GeV
  \item[$\bullet$]
    $P_e=P_{\bar{e}}=-1,~P_t = P_{\tilde{t}}=0,~
     P_\gamma = P_{\tilde{\gamma}}=1$, \ \
     $N_{\ell}\simeq 2.4\times 10^4 \epsilon_{\ell}$\\
     ${\mit\Delta} \alpha_{h1}=0.023/\sqrt{\epsilon_{\ell}}$,~
     ${\mit\Delta} \alpha_{d}=0.015/\sqrt{\epsilon_{\ell}}$.
\hfill (\arabic{QQ})\addtocounter{QQ}{1}
 \item[$\bullet$]
    $P_e=P_{\bar{e}}=-1,~P_t = P_{\tilde{t}}=0,~
     P_\gamma = P_{\tilde{\gamma}}=1$, \ \
     $N_{\ell}\simeq 2.4\times 10^4 \epsilon_{\ell}$\\
     ${\mit\Delta} \alpha_{h2}=0.030/\sqrt{\epsilon_{\ell}}$,~
     ${\mit\Delta} \alpha_{d}=0.015/\sqrt{\epsilon_{\ell}}$.
\hfill (\arabic{QQ})\addtocounter{QQ}{1}
\end{description}
\item[] $\oplus$ Final bottom-quark detection
\begin{description}
  \item[] $m_H=100$ GeV
  \item[$\bullet$]
    $P_e=P_{\bar{e}}=-1,~P_t = P_{\tilde{t}}=1/\sqrt{2},~
     P_\gamma = - P_{\tilde{\gamma}}=-1/\sqrt{2}$, \ \
     $N_{b}\simeq 4.2\times 10^4 \epsilon_{b}$\\
     ${\mit\Delta} \alpha_{h1}=0.058/\sqrt{\epsilon_{b}}$,~
     ${\mit\Delta} \alpha_{d}=0.026/\sqrt{\epsilon_{b}}$.
\hfill (\arabic{QQ})\addtocounter{QQ}{1}
  \item[] $m_H=300$ GeV
  \item[$\bullet$]
    $P_e=P_{\bar{e}}=-1,~P_t = P_{\tilde{t}}=1/\sqrt{2},~
     P_\gamma = -P_{\tilde{\gamma}}=-1/\sqrt{2}, \ \
     N_{b}\simeq 4.2\times 10^4 \epsilon_{b}$\\
     ${\mit\Delta} \alpha_{h1}=0.009/\sqrt{\epsilon_{b}}$,~
     ${\mit\Delta} \alpha_{h2}=0.074/\sqrt{\epsilon_{b}}$.
\hfill (\arabic{QQ})\addtocounter{QQ}{1}
  \item[$\bullet$]
    $P_e=P_{\bar{e}}=1,~P_t = P_{\tilde{t}}=1/\sqrt{2},~
     P_\gamma = -P_{\tilde{\gamma}}=-1/\sqrt{2}$, \ \
     $N_{b}\simeq 4.2\times 10^4 \epsilon_{b}$\\
     ${\mit\Delta} \alpha_{h1}=0.025/\sqrt{\epsilon_{b}}$,~
     ${\mit\Delta} \alpha_{d}=0.019/\sqrt{\epsilon_{b}}$.
\hfill (\arabic{QQ})\addtocounter{QQ}{1}
 \item[$\bullet$]
    $P_e=P_{\bar{e}}=1,~P_t = P_{\tilde{t}}=1/\sqrt{2},~
     P_\gamma = -P_{\tilde{\gamma}}=1/\sqrt{2}$, \ \
     $N_{b}\simeq 4.2\times 10^4 \epsilon_{b}$\\
     ${\mit\Delta} \alpha_{h2}=0.065/\sqrt{\epsilon_{b}}$,~
     ${\mit\Delta} \alpha_{d}=0.010/\sqrt{\epsilon_{b}}$.
\hfill (\arabic{QQ})\addtocounter{QQ}{1}
  \item[] $m_H=500$ GeV
  \item[$\bullet$]
    $P_e=P_{\bar{e}}=-1,~P_t = P_{\tilde{t}}=1,~
     P_\gamma = P_{\tilde{\gamma}}=0$, \ \
     $N_{b}\simeq 4.6\times 10^4 \epsilon_{b}$\\
     ${\mit\Delta} \alpha_{h1}=0.030/\sqrt{\epsilon_{b}}$,~
     ${\mit\Delta} \alpha_{d}=0.018/\sqrt{\epsilon_{b}}$.
\hfill (\arabic{QQ})\addtocounter{QQ}{1}
 \item[$\bullet$]
    $P_e=P_{\bar{e}}=-1,~P_t = P_{\tilde{t}}=1,~
     P_\gamma = P_{\tilde{\gamma}}=0$, \ \
     $N_{b}\simeq 4.6\times 10^4 \epsilon_{b}$\\
     ${\mit\Delta} \alpha_{h2}=0.028/\sqrt{\epsilon_{b}}$,~
     ${\mit\Delta} \alpha_{d}=0.014/\sqrt{\epsilon_{b}}$. 
\hfill (\arabic{QQ})\setcounter{QQE}{\arabic{QQ}}
\end{description}
\end{description}
\setcounter{equation}{\arabic{QQ}}
Using these results one can find (given $m_H$) the most
suitable polarization for a determination of a given pair of
coefficients.

Note that it is difficult to simultaneously determine $\alpha_{\gamma 1}$
and $\alpha_{\gamma 2}$ in either the two- or three-parameter
analyses. Also, although we did find some new stable solutions
that would allow for a determination of
$\alpha_{\gamma 1}$ in the lepton analysis, the expected
precision is rather low. Nevertheless this demonstrates that
the use of purely linear laser polarization  is crucial
for measuring $\alpha_{\gamma 1}$. Unfortunately,
the statistical uncertainty for $\alpha_{\gamma 2}$ is still
large, even in this improved analysis, so we did not include it
among our examples. Other processes must be used to determine
this parameter; for a review see \cite{Atwood:2000tu}.

We found that there are many combinations of polarization
parameters that make uncertainties of $\alpha_{h1,h2}$ and
$\alpha_{d}$ relatively small. For instance, analyzing the
$b$-quark final state with the choices
    $P_e=P_{\bar{e}}=-1$, $P_t = P_{\tilde{t}}=1/\sqrt{2}$,
    $P_\gamma = -P_{\tilde{\gamma}}=-1/\sqrt{2}$
enables us to probe the Higgs-photon couplings
$\alpha_{h1}$ and $\alpha_{h2}$ of a
$ 300$~GeV Higgs-boson.

As already mentioned, the results are obtained for
${\mit\Lambda}=$ 1 TeV. If the scale of new physics 
equals ${\mit\Lambda}=\lambda$ TeV, then all the above
results (${\mit\Delta}\alpha_i$) are replaced with
${\mit\Delta}\alpha_i/\lambda^2$, which means that the right-hand
sides of eqs.(\arabic{QQF})--(\arabic{QQE}) giving
${\mit\Delta}\alpha_i$ are all multiplied by $\lambda^2$.

Some additional comments are in order here.
\bit
\item 
If we are only interested in  measuring the decay coefficient $\alpha_d$,
then the optimal polarization should be adjusted to maximize the
top-production with no significant Higgs exchange contribution (this is
because we keep only linear terms in the anomalous couplings).
However, if $\alpha_d$ and $\alpha_{h1}$ or $\alpha_{h2}$ are to
be determined, then certain compromise of the SM $t\bar{t}$-production
rate is necessary as one also needs a significant contribution
from the Higgs-boson exchange.
\item 
If, on the other hand, only Higgs couplings are to be measured,
then the optimal polarization would make the Higgs-exchange
diagram as large as possible. It is obvious that for the most
precise determination of the $\gaga H$ couplings, one should
go to the resonance region\footnote{That would require adjustments
    of polarizations of the initial electron and laser beams,
    tuning initial electron energies and choosing large conversion
    distance, for details see \cite{Borden:1993cw}. Then the
    $\gaga$ spectrum would peak at $\sqrt{s_{\gaga}}\simeq
    0.8 \sqrt{s_{e\bar{e}}}$. Here, since we do not consider
    $m_H=400$ GeV, we are never in the resonance region, as
    mentioned in Introduction.}
in order to increase the Higgs
production rate. A detailed study of $C\!P$-violating effects
in $\gaga \to H$ has been performed, e.g., in \cite{Gounaris:1997ef}.
There, for the luminosity $L_{e\bar e} = 20$~fb$^{-1}$, the authors
estimate 3-$\sigma$ limits for $\alpha_{h2}$
($d_{\gaga}=(v/{\mit\Lambda})^2\alpha_{h2}+\cdots$ in the notation
of \cite{Gounaris:1997ef}) at the level of $10^{-3}$--$10^{-4}$
depending on the Higgs-boson mass. Correcting for the luminosity
adopted here ($L_{e\bar e} = 500$~fb$^{-1}$) it corresponds to
our 1-$\sigma$ uncertainty for $\alpha_{h2}$ also of the order
of $10^{-3}$--$10^{-4}$, so smaller by about two orders of
magnitude than the precision obtained here for the off-resonance
region.
\eit

\sec{Comparing $\mib{e}\bar{\mib{e}}$ and $\mib{\gamma\gamma}$
colliders}

Since both $t\bar{t}\gamma$ and $tbW$ couplings contribute to
$\gamma\gamma\to t\bar{t}$ and $e\bar{e}\to t\bar{t}$,
it is pertinent to compare the sensitivity to those
anomalous couplings in these two types of colliders.

\bit
\item $\mib{e\bar{e}}$ {\bf colliders} \\
The assumption that the on-mass-shell $\ttbar$ are produced through
$s$-channel {(axial-)}vector-boson-exchange fixes all the kinematics
in $\ttbar Z$ and $\ttbar \gamma$ vertices as a function of
$\sqrt{s}$ and $\mt$ only. For the subsequent two-body on-shell
$t$ decay, the kinematics is also fixed (just by masses).
Therefore, in this framework, we can perform a very general
analysis without worrying about the momentum
dependence of all the effective vertices, i.e., not referring
to the effective Lagrangian but treating the anomalous couplings
as form factors (which could be momentum dependent). As we have
shown earlier \cite{Grzadkowski:1996kn,Grzadkowski:2000nx},
momentum distributions of the secondary lepton and
the $b$-quark can serve as a mean to measure of {\it real parts}
of the anomalous form factors. There, we used the anomalous magnetic-
and electric-dipole-type couplings $\delta C_{\gamma}$ and
$\delta\!D_{\gamma}$ for $t\bar{t}\gamma$ vertex and $f_2^R$ for
$tbW$ vertex given; see appendices A2 and A3.
The correspondence to $\alpha_{\gamma 1}$,
$\alpha_{\gamma 2}$ and $\alpha_d$ is
\begin{equation}
\delta C_{\gamma}=-\frac{4\sqrt{2}m_t v}{g{\mit\Lambda}^2}
\alpha_{\gamma 1},\ \ \
\delta\!D_{\gamma}=i\frac{4\sqrt{2} m_t v}{g{\mit\Lambda}^2}
\alpha_{\gamma 2}, \ \ \
{\rm Re}(f_2^R)=\alpha_d,~\label{relation}
\end{equation}
where $g$ is the $SU(2)$ coupling and $v$ is the electroweak vacuum
expectation value ($\simeq 250$ GeV). Within the effective-Lagrangian
framework $\alpha_{\gamma 1,\gamma 2}$ are real numbers,
so $\delta\!D_{\gamma}$ is purely imaginary. Since only the real
parts of the form factors can be measured through the
distributions of the final fermions, $e\bar e$ colliders
are sensitive only to $\delta C_{\gamma}$ ($\alpha_{\gamma 1}$)
and $f_2^R$ ($\alpha_d$).\footnote{Note that we are discussing
    only the couplings which contribute to {\em both} $e\bar{e}$ and
    $\gamma\gamma$ processes. Of course, at $e\bar{e}$ colliders it is
    possible to determine the $t\bar{t} Z$ anomalous couplings
    to which $\gamma\gamma$ machines are completely blind; see
    \cite{Grzadkowski:1996kn,Grzadkowski:2000nx}.}

\item $\mib{\gamma\gamma}$ {\bf colliders} \\
Due to the presence of the $t$-channel diagram, in the case of
the $\gaga$ collider the kinematics of the $\ttbar \gamma$ vertex
is not fixed by $\sqrt{s}$ and the masses. In order to calculate
distributions of secondary particles, one would need to integrate
over momenta on which the $\ttbar \gamma$ form factor may depend.
Therefore for $\gaga$ colliders we will not go  beyond the
effective-Lagrangian framework in which all the anomalous couplings
are just given by constant coefficients.
In \cite{Grzadkowski:2003tf} we have
shown that for the $\gaga$ scattering we could in general
determine {\it both} real {\it and} imaginary parts of the
anomalous $\gamma$ couplings.\footnote{Indeed, $\alpha_{\gamma 1}$
    and $\alpha_{\gamma 2}$ are respectively the real part and
    the imaginary part of one parameter as shown in (\ref{Ag1})
    and (\ref{Ag2}) in appendix A1. Calculating cross sections
    within our approximation,
    the imaginary part of any coupling cannot contribute unless the
    Levi-Civita tensor terms appearing in $\gamma$-matrix calculations
    survive. In case of $e\bar{e}\to t\bar{t} \to \ell X$ process,
    we do not have enough number of independent vectors to keep those
    terms non-vanishing, while we do have for $\gamma\gamma$ process.
    In order to keep the Levi-Civita tensor terms non-zero in the
    final result in $e\bar{e}$-process analyses, we would need to
    introduce some additional independent vectors by, e.g., defining
    an angular asymmetry, see for instance \cite{Chang:1993fu}.}
\eit

Because of the above remarks, in order to compare $e\bar{e}$
and $\gaga$ colliders we will adopt the framework of the
effective Lagrangian. Then it is clear that there are only two
couplings which can be measured at both machines:
$\alpha_{\gamma 1}$ and $\alpha_d$. Therefore, we present
results of one-parameter OO analysis (i.e., assuming that only one
coupling is to be determined at a time) for them using the
final-lepton distributions. We show the highest expected precision
of each parameter obtained while varying the polarization
parameters.

\noindent
$\bullet\ e\bar{e} \to t\bar{t} \to \ell X$

$e$-1) ${\mit\Delta}\alpha_{\gamma 1}
=0.02/\sqrt{\epsilon_{\ell}}$\ \ \
for\ \ $P_e=-1$ and $P_{\bar{e}}=+1$,\ \
$N_{\ell}\simeq 1.5\times 10^5 \epsilon_{\ell}$

$e$-2) ${\mit\Delta}{\rm Re}(f_2^R)
=0.003/\sqrt{\epsilon_{\ell}}$\ \ \
for\ \ $P_e=-1$ and $P_{\bar{e}}=+1$,\ \
$N_{\ell}\simeq 1.5\times 10^5 \epsilon_{\ell}$

\noindent
$\bullet\ \gamma\gamma \to t\bar{t} \to \ell X$

$\gamma$-1) ${\mit\Delta}\alpha_{\gamma 1}
=0.03/\sqrt{\epsilon_{\ell}}$\ \ \
for\ \ $P_e=P_{\bar{e}}=\pm 1$ and $P_t=P_{\tilde{t}}=1$,
\ \
$N_{\ell}\simeq 1.0\times 10^4 \epsilon_{\ell}$

$\gamma$-2) ${\mit\Delta}\alpha_d
=0.01/\sqrt{\epsilon_{\ell}}$\ \ \
for\ \ $P_e=P_{\bar{e}}=-1$ and
$P_{\gamma}=P_{\tilde{\gamma}}=1$,
\ \
$N_{\ell}\simeq 2.4\times 10^4 \epsilon_{\ell}$

\medskip

As one can see the precision obtained for $\delta C_{\gamma}$
($\alpha_{\gamma 1}$) is of the same order although the
$e\bar{e}$ machine seems to be slightly favored. On the other
hand the precision for $f_2^R$ ($\alpha_d$) is much better
in $e\bar{e}$ than in $\gamma\gamma$. This simply comes
from the difference in the expected event numbers $N_{\ell}$
obtained for each optimal polarizations for the same $e\bar{e}$
luminosity. So, the $e\bar{e}$ machine is superior as far as
the determination of the
top-quark decay parameters is concerned.

\sec{Conclusions and discussions}

We have performed a detailed analysis of the process $\gamma\gamma
\to t\bar{t}\to \ell X/b X$ in order to find optimal beam
polarizations that minimize uncertainties in the determination
of $t\bar{t}\gamma$- , $tbW$- and $\gamma\gamma H$-coupling
parameters. To estimate the uncertainties we have applied
the optimal-observable procedure to the
final lepton/$b$-quark momentum distribution in $\gamma\gamma
\to t\bar{t}\to \ell X/b X$. We have also compared the $e\bar{e}$
and $\gamma\gamma$ colliders from the point of view of the
anomalous-top-quark-coupling determination.

Applying the optimal observable technique, we have again encountered
the problem of ``unstable-solutions" (see also
\cite{Grzadkowski:2003tf}) and concluded that there is no
stable solution when trying to determine more than
three anomalous couplings simultaneously. However, in contrast to
\cite{Grzadkowski:2003tf}, allowing for more polarization choices,
we have obtained stable solutions with three couplings.
We also found a number of two-parameter solutions, most of
which allow for the determination of the
$\gaga H$ and $tbW$ couplings.
The expected precision of the measurement of the Higgs coupling
is of the order of $10^{-2}$ (for the scale of new physics
${\mit\Lambda}=1$ TeV). This shows that the $\gamma\gamma$
collider will be useful for testing the Higgs
sector of the SM.

We also found that $e\bar{e}$ colliders will do
slightly better than $\gaga$ colliders for
the determination of $C\!P$-conserving
$t\bar{t}\gamma$ and $tbW$ couplings
(assuming the validity of the effective-Lagrangian framework).
One should not forget, however, that $e\bar{e}$ colliders
can only measure the real part of the $t\bar{t}\gamma$ and $tbW$
couplings as long as we perform full integration over the
final-particle momenta.

Apart from the $t\bar{t}\gamma$- and $tbW$-coupling determinations,
the $\gamma\gamma\to t\bar{t}$ and $e\bar{e}\to t\bar{t}$
processes are sensitive to different types of couplings. The
former provides information on $\gaga H$ couplings, while
the $t\bar{t} Z$ couplings can be tested only via the latter.
Therefore it is fair to conclude that the measurements
from both colliders will complement one another.
In this respect it should be noted that $\gaga H$ coupling could also be
measured at $e\bar e$ colliders using final states such as
 $e\bar e \to \gamma H$
\cite{Hagiwara:1993sw}. However the expected uncertainty
is two orders of magnitude larger than at $\gaga$ colliders,
see \cite{Gounaris:1997ef,Hagiwara:1993sw}. Therefore, the
$\gaga$ collider is definitely superior as far as the
determination of $\gamma\gamma H$ couplings is concerned,

Let us consider the top-quark-coupling determination in an
ideal case such that the beam parameters could be easily
tuned and that the energy is sufficient for the on-shell
Higgs-boson production, assuming that the Higgs-boson
mass is known from the Large Hadron Collider. Then the
best strategy would be to adjust polarizations and tune
the initial electron energies to construct
semi-monochromatic $\gaga$ beams such that
$\sqrt{s_{\gaga}}\simeq m_H$ and on-shell Higgs bosons are
produced. This would allow for precise $\alpha_{h1,h2}$
measurement, so the virtual Higgs effects in $\gaga\to\ttbar$
would be calculable. Unfortunately, as we have shown earlier,
it is difficult to measure $\alpha_{\gamma2}$ by looking
just at $\ell X/b X$ final states from $\gaga\to\ttbar$.
Therefore to fix $\alpha_{\gamma2}$, one should, for example, measure
the asymmetries described in \cite{Choi:1995kp} to determine
the top-quark electric-dipole moment, which is proportional
to $\alpha_{\gamma2}$.
Then, following the analysis presented here,
one can determine $\alpha_{\gamma1}$ and $\alpha_d$.

Finally, one must not forget that it is necessary to take into
account carefully the Standard Model contribution with radiative
corrections when trying to determine the anomalous couplings
in a fully realistic analyses. In particular this is significant
when we are interested in $C\!P$-conserving couplings since
the SM contributions there are not suppressed unlike the
$C\!P$-violating terms. On this subject, see for instance
\cite{Brandenburg:2005uu}.

\vspace{0.6cm}
\centerline{ACKNOWLEDGMENTS}

\vspace{0.3cm}
This work is supported in part by the State Committee for
Scientific Research (Poland) under grant 1~P03B~078~26 in
the period 2004--2006, the Grant-in-Aid for Scientific
Research No.13135219 and No.16540258 from the Japan
Society for the Promotion of Science, and the Grant-in-Aid
for Young Scientists No. 17740157 from the Ministry of
Education, Culture, Sports, Science and Technology of Japan.

\vskip 1.0cm
\centerline{APPENDIX}

\vspace*{0.3cm}
\noindent \hskip -0.46cm 
{\bf A1. Dimension-6 operators inducing {\boldmath $\gamma\gamma \to 
\ttbar$}}

\noindent
Following the Buchm\"uller and Wyler scenario \cite{Buchmuller:1986jz},
operators  of dim.6 that could contribute to the continuum
top-quark-production process $\pptt$ read:
\begin{equation}
\begin{array}{ll}
\ocal_{uB}'=(\bar{q}\sigma^{\mu\nu}u)\tilde{\varphi} B_{\mu\nu},
& \ocal_{uW}
=(\bar{q}\sigma^{\mu\nu}\tau^i u)\tilde{\varphi} W_{i\;\mu\nu}.
\end{array}
\label{prod-cont-eff}
\end{equation}
Each of the above operators contains both $C\!P$-violating and
$C\!P$-conserving parts.

On the other hand, the following operators contribute to $\pptt$
through the resonant $s$-channel Higgs-boson exchange:
\begin{equation}
\begin{array}{ll}
\ocal_{\varphi\tilde{W}}
=(\varphi^\dagger \varphi)\tilde{W}_{\mu\nu}^i W^{i\; \mu\nu}, &
\ocal_{\varphi W}
=(\varphi^\dagger \varphi)W_{\mu\nu}^i W^{i\; \mu\nu}/2, \\
\ocal_{\varphi\tilde{B}}
=(\varphi^\dagger \varphi)\tilde{B}_{\mu\nu} B^{\mu\nu}, &
\ocal_{\varphi B}
=(\varphi^\dagger \varphi)B_{\mu\nu} B^{\mu\nu}/2, \\
\ocal_{\tilde{W}B}
=(\varphi^\dagger\tau^i \varphi)\tilde{W}_{\mu\nu}^i B^{\mu\nu}, &
\ocal_{WB}
=(\varphi^\dagger\tau^i \varphi)W_{\mu\nu}^i B^{\mu\nu}.
\end{array}
\label{prod-res}
\end{equation}
The operators that contain the dual tensors
(e.g., $\tilde{B}_{\mu\nu}
\equiv \epsilon_{\mu\nu\alpha\beta}B^{\alpha\beta}/2$ with
$\epsilon_{0123}=+1$) are $C\!P$ odd while the remaining are
$C\!P$ even.

All these operators lead to the following Feynman rules for on-shell
photons, which are necessary for our calculations: \\ \\
(1) $C\!P$-conserving $t\bar{t}\gamma$ vertex
\begin{equation}
\frac{\sqrt{2}}{{\mit\Lambda}^2}v \alpha_{\gamma 1}\,
\slak\gamma_\mu,
\end{equation}
(2) $C\!P$-violating $t\bar{t}\gamma$ vertex
\begin{equation}
i\frac{\sqrt{2}}{{\mit\Lambda}^2}v \alpha_{\gamma 2}\,
\slak\gamma_\mu \gamma_5, \label{cpv-vertex}
\end{equation}
(3) $C\!P$-conserving $\gamma\gamma H$ vertex
\begin{eqnarray}
&&-\frac{4}{{\mit\Lambda}^2}v \alpha_{h1}\,
\bigl[\:
(k_1 k_2)g_{\mu\nu}-k_{1\nu}k_{2\mu}
\:\bigr],
\end{eqnarray}
(4) $C\!P$-violating $\gamma\gamma H$ vertex
\begin{eqnarray}
&&\frac{8}{{\mit\Lambda}^2}v \alpha_{h2}\,
k_1^\rho k_2^\sigma \epsilon_{\rho\sigma\mu\nu},
\end{eqnarray}
where $k$ and $k_{1,2}$ are incoming photon momenta,
$v$ is the EW vacuum
expectation value ($\simeq 250$ GeV)
and
$\alpha_{\gamma 1,\gamma 2,h1,h2}$ are defined as
\begin{eqnarray}
&&\alpha_{\gamma 1}\equiv
\sin\theta_W{\rm Re}(\alpha_{uW})
+\cos\theta_W{\rm Re}(\alpha'_{uB}), \label{Ag1}
\\
&&\alpha_{\gamma 2}\equiv
\sin\theta_W{\rm Im}(\alpha_{uW})
+\cos\theta_W{\rm Im}(\alpha'_{uB}), \label{Ag2}
\\
&&\alpha_{h 1}\equiv
\sin^2\theta_W{\rm Re}(\alpha_{\varphi W})
+\cos^2\theta_W{\rm Re}(\alpha_{\varphi B}) 
-2\sin\theta_W \cos\theta_W{\rm Re}(\alpha_{WB}),
\\
&&\alpha_{h 2}\equiv
\sin^2\theta_W{\rm Re}(\alpha_{\varphi \tilde{W}})
+\cos^2\theta_W{\rm Re}(\alpha_{\varphi \tilde{B}}) 
-\sin\theta_W \cos\theta_W{\rm Re}(\alpha_{\tilde{W}B}).
\end{eqnarray}
In our notation, the SM $f\bar{f}\gamma$ coupling
is given by
\[
eQ_f \gamma_\mu,
\]
where $e$ is the proton charge and $Q_f$ is $f$'s electric charge
in unit of $e$ (e.g., $Q_u = 2/3$).

\vskip 0.4cm
\noindent \hskip -0.46cm 
{\bf A2. Dimension-6 operators inducing {\boldmath $t\to bW$}}

\noindent
The top-quark decay vertex is also affected by some dim.6 operators.
For the on-mass-shell $W$ boson it will be sufficient to consider
just the following contributions to the
$tbW$ amplitude since other possible terms do
not interfere with the SM tree-level vertex when $m_b$ is neglected:
\begin{equation}
{\mit\Gamma}^{\mu}_{tbW}=-{g\over\sqrt{2}}\:
\bar{u}(p_b)\biggl[\,\gamma^{\mu} f_1^L P_L
-{{i\sigma^{\mu\nu}k_{\nu}}\over M_W}
f_2^R P_R\,\biggr]u(p_t),\label{ffdef}\\
\end{equation}
where $g$ denotes the $SU(2)$ gauge coupling constant, 
$P_{L,R}\equiv (1\mp \gamma_5)/2$, and $f_1^L$ and $f_2^R$
are given by
\begin{eqnarray}
&&f^L_1=1+\frac{v}{{\mit\Lambda}^2}
  \Bigl[\:
\frac{m_t}{2}(\alpha_{Du}-\alpha_{\bar{D}u})
  -2v\alpha_{\varphi q}^{(3)}\:\Bigr], \label{f1L} \\
&&f^R_2=-\frac{v}{{\mit\Lambda}^2}M_W
  \Bigl[\:\frac{4}{g}\alpha_{uW}
  +\frac12(\alpha_{Du}-\alpha_{\bar{D}u})
\:\Bigr],  \label{f2R}
\end{eqnarray}
with $\alpha_{Du}$, $\alpha_{\bar{D}u}$ and $\alpha_{\varphi q}^{(3)}$
being correspondingly the coefficients of the following operators:
\beq
{\cal O}_{Du}=(\bar{q}D_{\mu}u)D^{\mu}\tilde{\varphi},\;\;\;\;
{\cal O}_{\bar{D}u}=(D_{\mu}\bar{q})u D^{\mu}\tilde{\varphi},\;\;\;\;
{\cal O}_{\varphi q}^{(3)}
=i(\varphi^{\dagger}D_{\mu}\tau^i\varphi)(\bar{q}\gamma^{\mu}\tau^i q).
\eeq
In the main text, we express ${\rm Re}(f_2^R)$ as $\alpha_d$.

\vskip 0.4cm
\noindent \hskip -0.46cm 
{\bf A3. General invariant amplitude of {\boldmath $e\bar{e}\to \ttbar$}}

\noindent
We assume that all
non-standard effects in the production process  $e\bar{e}\to\ttbar$ can be 
represented
by the following corrections to the photon and $Z$-boson vertices
contributing to the $s$-channel diagrams:
\begin{equation}
{\mit\Gamma}_{vt\bar{t}}^{\mu}=
\frac{g}{2}\,\bar{u}(p_t)\,\Bigl[\,\gamma^\mu \{A_v+\delta\!A_v
-(B_v+\delta\!B_v) \gamma_5 \}
+\frac{(p_t-p_{\bar{t}})^\mu}{2m_t}(\delta C_v-\delta\!D_v\gamma_5)
\,\Bigr]\,v(p_{\bar{t}}),\ \label{tt-amp}
\end{equation}
where $v=\gamma,Z$ and
\[
\av=\frac{4}{3}\sw,\ \ \bv=0,\ \ \az=\frac{v_t}{2\cw},\ \ \bz
=\frac{1}{2\cw}
\]
with $v_t = 1-8\sin^2 \theta_W/3$.
In addition, contributions to the vertex proportional to $(p_t +
p_{\bar{t}})^\mu$ are also allowed, but their effects vanish in the
limit of zero electron mass. Therefore, we can say that this form
is practically the most general invariant one.
Among the above new form factors, $\delta\!A_{\gamma,Z},
\delta\!B_{\gamma,Z}$ and $\delta C_{\gamma,Z}$
are parameterizing $C\!P$-conserving, while $\delta\!D_{\gamma,Z}$ describes
$C\!P$-violating non-standard interactions.
A complete list of these non-standard couplings expressed through
coefficients of dim.6 effective operators
is to be found, e.g., in the
second paper in \cite{Grzadkowski:1996kn}.

\vspace*{0.8cm}

\end{document}